\documentclass[iicol,pdflatex,sn-nature]{sn-jnl}

\usepackage{graphicx}%
\usepackage{multirow}%
\usepackage{amsmath,amssymb,amsfonts}%
\usepackage{amsthm}%
\usepackage{mathrsfs}%
\usepackage[title]{appendix}%
\usepackage{xcolor}%
\usepackage{textcomp}%
\usepackage{manyfoot}%
\usepackage{booktabs}%
\usepackage{algorithm}%
\usepackage{algorithmicx}%
\usepackage{algpseudocode}%
\usepackage{listings}%
\usepackage{hyperref}
\usepackage[capitalize]{cleveref}

\usepackage{makecell}

\crefname{section}{Sec.}{Secs.}
\Crefname{section}{Section}{Sections}
\Crefname{table}{Table}{Tables}
\crefname{table}{Table}{Tables}
\Crefname{figure}{Figure}{Figures}
\crefname{figure}{Fig.}{Figs.}
\Crefname{equation}{Equation}{Equations}
\crefname{equation}{Eq.}{Eqs.}
\crefname{algocf}{alg.}{algs.}
\Crefname{algocf}{Algorithm}{Algorithms}

\begin{document}

\onecolumn

\title[A generative framework]{A generative framework for unified representations bridging electrophysiological and haemodynamic brain activity}

\author[1]{\fnm{Weiheng} \sur{Yao}}

\author[1]{\fnm{Changhong} \sur{Jing}}

\author[1,2]{\fnm{Xuhang} \sur{Chen}}

%

\author[2]{\fnm{Chi-Man} \sur{Pun}}

\author[3]{\fnm{Michael Kwok-Po} \sur{Ng}}

\author*[1]{\fnm{Shuqiang} \sur{Wang}}\email{sq.wang@siat.ac.cn}

\affil[1]{\orgdiv{Shenzhen Institutes of Advanced Technology}, \orgname{Chinese Academy of Sciences}, \orgaddress{\city{Shenzhen}, \postcode{518055}, \state{Guangdong}, \country{China}}}


\affil[2]{\orgdiv{Department of Computer and Information Science}, \orgname{University of Macau}, \orgaddress{\city{Macau}, \postcode{999078}, \country{China}}}

%

\affil[3]{\orgdiv{Department of Mathematics}, \orgname{Hong Kong Baptist University}, \orgaddress{\city{Hong Kong}, \postcode{999077},  \country{China}}}

\abstract{
Electrophysiological and haemodynamic recordings resolve brain activity at complementary temporal and spatial scales, yet paired acquisition remains difficult to scale. We propose a generative framework that aligns paired multimodal recordings in a unified representation space and generates modality-specific haemodynamic responses from EEG. In paired cohorts, the framework generated fMRI BOLD and fNIRS haemoglobin responses that reproduced regional dynamics, cortical maps, functional connectivity and task structure. Generated BOLD followed Real BOLD time courses and cortical patterns, while Generated haemoglobin responses tracked task-evoked HbO and HbR dynamics. The framework enabled decoding across visual, motor-imagery and mental-arithmetic tasks, independent-record decoding, external BOLD localization and low-sample augmentation. Paired feature-space calibration supported bidirectional retrieval. The proposed framework extends cortical analysis and brain decoding to EEG recordings when Real haemodynamic data are sparse.
}

\maketitle
\twocolumn

\begin{figure*}[tp]
\centering
\includegraphics[width=\linewidth,height=0.80\textheight,keepaspectratio]{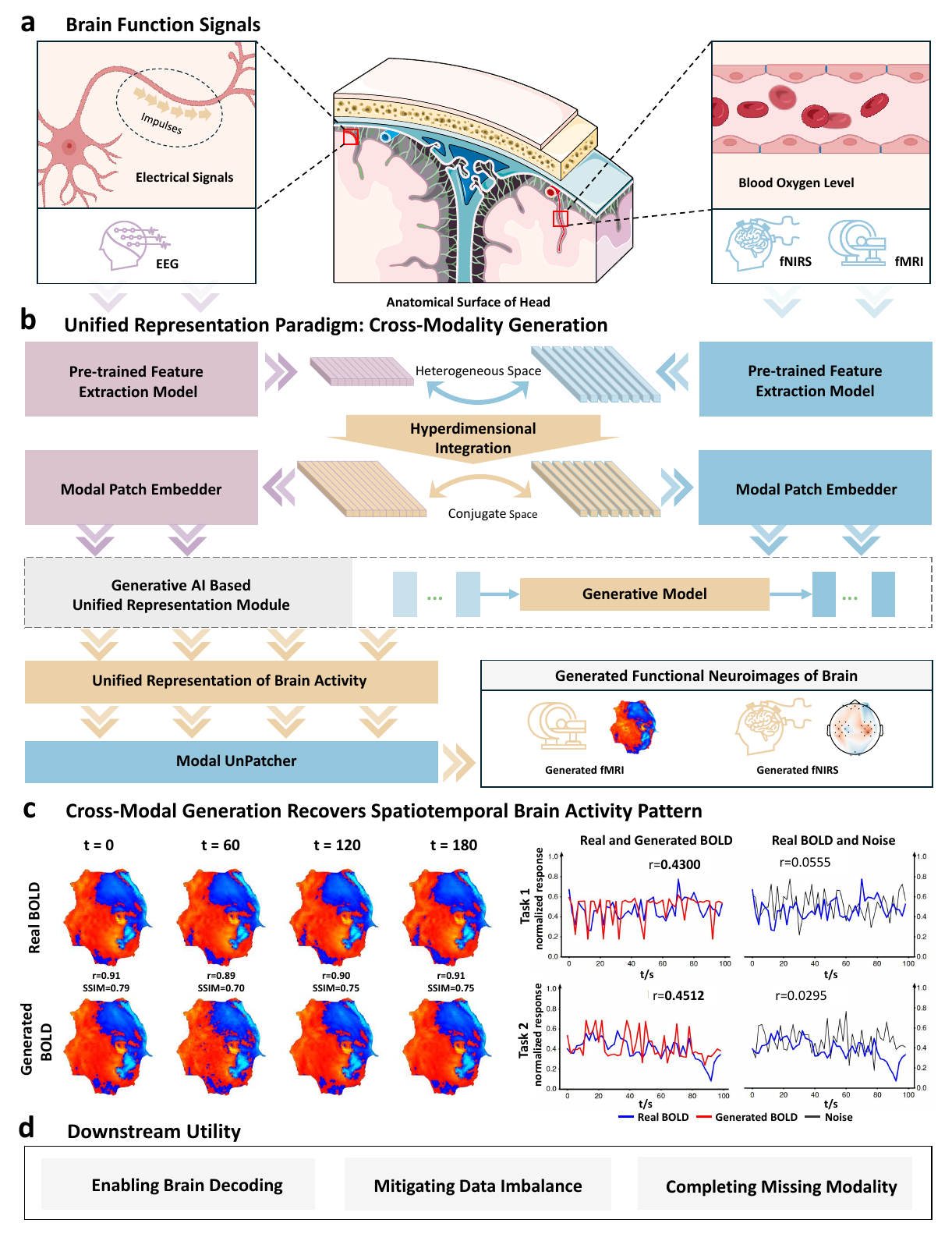}
\caption{\textbf{Generative framework for unified representations bridging electrophysiological and haemodynamic brain activity.} \textbf{a,} EEG records scalp electrical potentials, fMRI detects BOLD contrast through magnetic resonance, and fNIRS estimates HbO and HbR responses through optical absorption. The anatomical surface of the head was adapted from Servier Medical Art by Servier under the Creative Commons Attribution 3.0 Unported licence. \textbf{b,} Pre-trained feature extraction models and modal patch embedders project electrophysiological and haemodynamic inputs into heterogeneous spaces. Hyperdimensional integration aligns these spaces in a conjugate representation. A generative model within the unified representation module produces target-domain codes, and modal unpatchers project the codes back to fMRI BOLD and fNIRS haemoglobin outputs. \textbf{c,} Generated BOLD reproduces spatiotemporal cortical activity patterns and follows Real BOLD time courses more closely than noise controls. Spatial Pearson \(r\) and the structural similarity index measure cortical-map correspondence. \textbf{d,} Downstream applications include brain decoding, data-imbalance mitigation and missing-modality completion.}
\label{fig1}
\end{figure*}

Brain activity is accessible through electrophysiological and haemodynamic recordings \cite{science1,cell1,nature2,nature3,nmi1,nmi2}. Electroencephalography (EEG) captures rapid electrical activity at high temporal resolution and has been used for cognitive assessment, disease characterization, individualized craving prediction and automated detection of harmful brain activity \cite{Sun2024AdaptiveEEG,Wang2026MSAEEG,Li2026HeroinEEGFC,Sun2025VIPEEGNet}. Functional magnetic resonance imaging (fMRI) provides blood-oxygen-level-dependent (BOLD) responses with high spatial coverage \cite{natureBOLD,natrevphy}. Functional near-infrared spectroscopy (fNIRS) provides oxyhaemoglobin and deoxyhaemoglobin responses through optical haemodynamic sensing. Wearable fNIRS platforms enable dense sampling for precision functional neuroimaging \cite{Jounghani2025WearablefNIRS}. EEG, fMRI and fNIRS capture complementary manifestations of brain function through distinct biophysical channels. Task-evoked spatial maps, haemodynamic response dynamics, functional connectivity and decodable task structure provide observable endpoints for calibration across these recording streams \cite{Cortex-Align,nature4,nature5}.

Paired EEG-fMRI and EEG-fNIRS recordings provide direct supervision for cross-domain calibration \cite{naturefMRI,nc2,nc4,jing1}. A generative model can convert this calibration into reusable haemodynamic outputs for analysing response dynamics, spatial maps, network relations and task structure.

Cross-modal haemodynamic generation gains scientific value when its outputs preserve the regional dynamics, functional connectivity and task-dependent structure used in neuroimaging analysis. These properties determine whether generated responses can support analyses normally performed with Real haemodynamic data.

Multimodal representation learning links source and target domains \cite{Baltrusaitis2017}. Pre-trained feature models and conditional generators provide the computational basis for learning this relation \cite{natmethods1,natmethods2,DiT_peebles2022scalable,GMAI_moor2023foundation,jing7,jing13,jing4,nmi4,nmi5}. Recent EEG-to-fMRI methods such as CATD have advanced latent alignment and diffusion-based generation \cite{CATD2025}. In this setting, EEG-to-haemodynamic generation is formulated in the target domain, with each output expressed in the coordinates used for BOLD or haemoglobin analysis. The same formulation supports missing-modality completion when an acquisition session contains EEG without a paired haemodynamic recording.

We propose a generative framework that aligns paired multimodal recordings in a unified representation space and generates BOLD responses in fMRI and haemoglobin responses in fNIRS (Fig. 1). The method combines feature extraction, modal patch embedding, conjugate alignment and target-domain generation with modality-specific calibration. Generated responses retained temporal, spatial and task-related structure across both haemodynamic domains and supported cortical mapping, functional-connectivity analysis, brain decoding and low-sample data augmentation.

\section*{Results}

\subsection*{The Framework Generates BOLD With Regional Response Dynamics}

\begin{figure*}[tp]
\centering
\includegraphics[width=\linewidth,height=0.90\textheight,keepaspectratio]{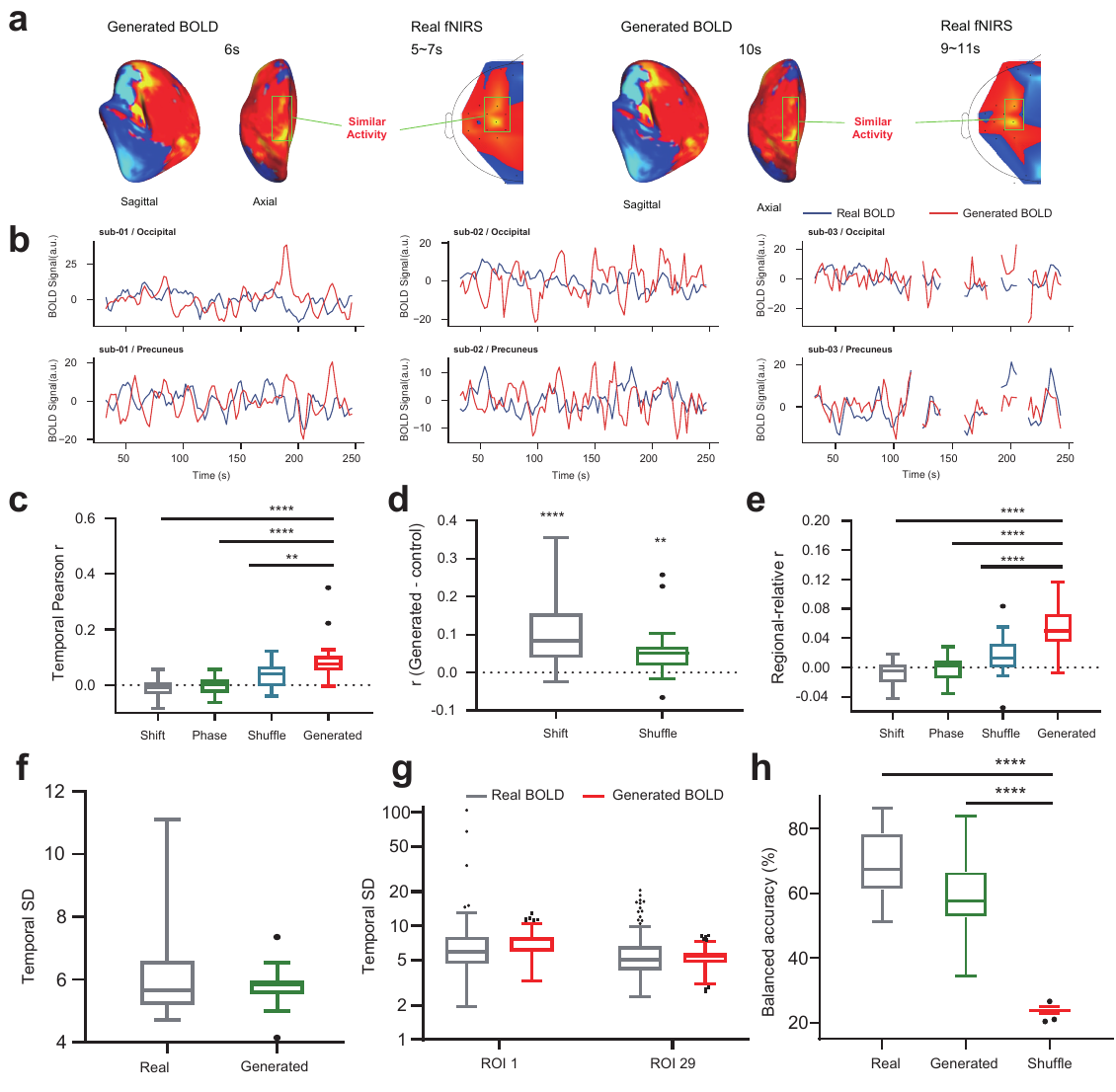}
\caption{\textbf{BOLD generation captures response dynamics and task structure.} \textbf{a,} Generated BOLD maps at 6 and 10 s are displayed alongside Real fNIRS haemoglobin response fields averaged over 5 to 7 and 9 to 11 s, respectively. \textbf{b,} Real BOLD and Generated BOLD time courses are shown in occipital and precuneus regions from three participants. \textbf{c,} Temporal Pearson $r$ compares Generated BOLD with shifted-input, phase-randomized and shuffled-pair controls. \textbf{d,} Generated minus control differences quantify gains attributable to the calibrated source-target pairing. \textbf{e,} Regional-relative $r$ evaluates structure beyond a shared global component. \textbf{f,} Whole-signal temporal standard deviation compares Real BOLD and Generated BOLD. \textbf{g,} ROI-level temporal standard deviation compares Real BOLD and Generated BOLD in two representative regions. \textbf{h,} Balanced accuracy compares Real BOLD features, Generated BOLD features and shuffled BOLD features in the same decoding analysis. Panels c to f and h summarize 21 participants, and panel g summarizes 141 recordings. Boxes show median, interquartile range and 1.5-IQR whiskers. Two-sided participant-level sign-flip tests with Holm correction were used for paired contrasts. Stars denote corrected significance levels with $^{*}P<0.05$, $^{**}P<0.01$, $^{***}P<0.001$ and $^{****}P<0.0001$.}
\label{fig2}
\end{figure*}

The BOLD target head generated responses with temporal and spatial structure. Spatial maps showed regional correspondence with Real fNIRS haemoglobin response fields at the displayed time windows, and participant-level traces tracked Real BOLD fluctuations in occipital and precuneus regions (Fig. 2a,b). Temporal Pearson $r$ was higher for Generated BOLD than for shifted-input, phase-randomized and shuffled-pair controls (Fig. 2c). Participant-level Generated minus control differences confirmed the advantage of calibrated source-target pairing over shifted inputs and shuffled pairs (Fig. 2d). Regional-relative $r$ remained positive after removal of a shared global component and exceeded all three controls (Fig. 2e). The framework therefore captured regional temporal variation alongside the shared component of the cortical response.

Real BOLD and Generated BOLD occupied overlapping temporal-variability ranges at the whole-signal and ROI levels (Fig. 2f,g). The similar central values in both representative regions showed that the BOLD target head maintained a response scale comparable to Real BOLD. Generated BOLD also achieved balanced accuracy above shuffled BOLD in the matched decoding analysis (Fig. 2h). This decoding advantage linked the reconstructed dynamics to task-related differences in the generated signals.

\subsection*{The BOLD Branch Generates Cortical Maps and Functional Connectivity}

\begin{figure*}[tp]
\centering
\includegraphics[width=\linewidth,height=0.86\textheight,keepaspectratio]{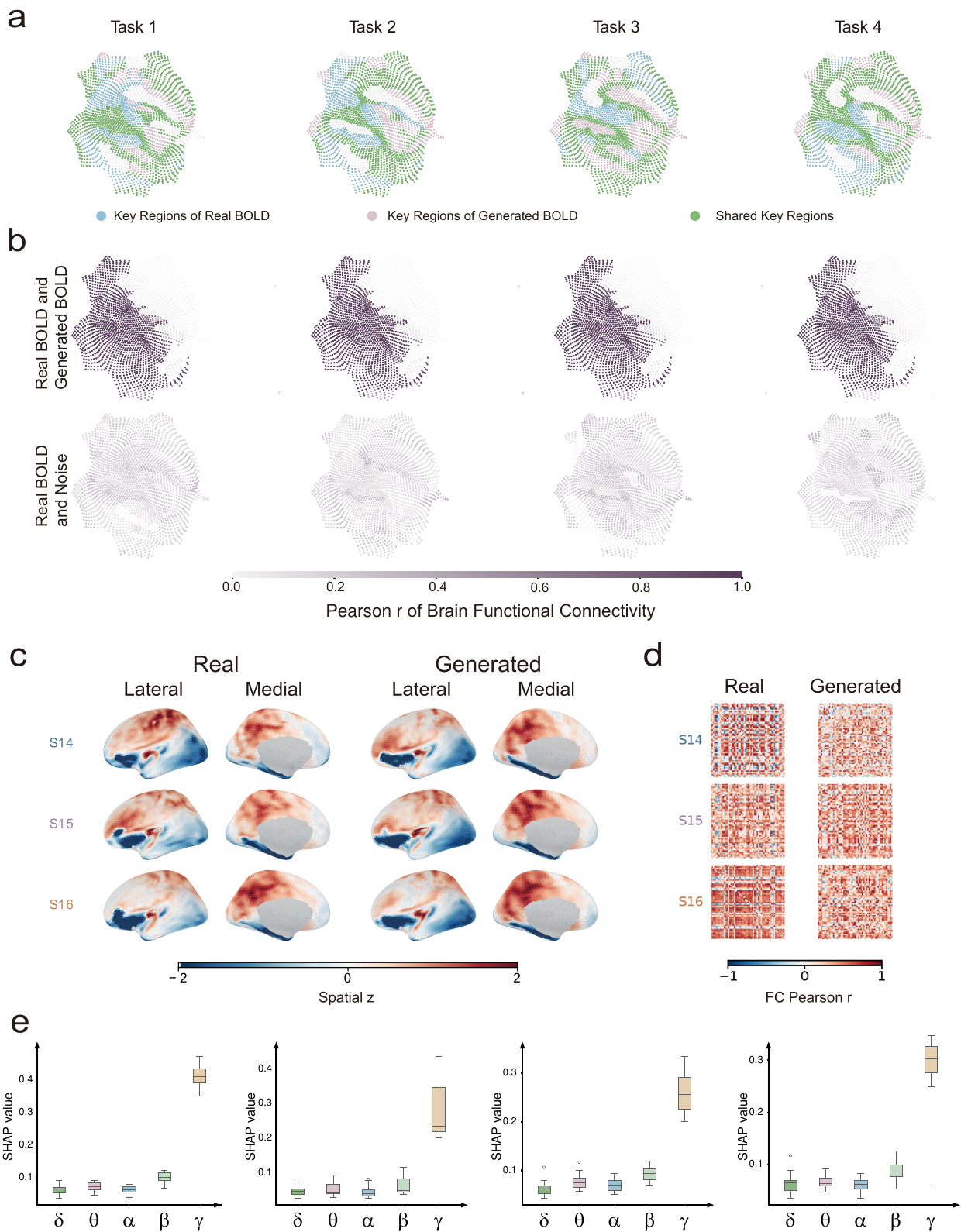}
\caption{\textbf{Generated BOLD supports cortical mapping and connectome analysis.} \textbf{a,} Key regions from Real BOLD maps, Generated BOLD maps and their shared overlap are shown for four tasks. \textbf{b,} Cortical functional-connectivity profiles compare Real-Generated correspondence with the Real-noise reference. \textbf{c,} Real BOLD cortical activity maps and Generated BOLD cortical activity maps are displayed in lateral and medial views for three participants. \textbf{d,} Real BOLD functional-connectivity matrices and Generated BOLD functional-connectivity matrices are shown for the same participants. \textbf{e,} SHAP values summarize frequency-band contributions to visual-task decoding from band-conditioned Generated BOLD. Boxes show median, interquartile range and 1.5-IQR whiskers.}
\label{fig3}
\end{figure*}

Generated BOLD was projected onto cortical anatomy. Key-region maps from Real BOLD and Generated BOLD showed task-wise overlap across the four visual conditions (Fig. 3a). The shared regions linked the generated responses to the spatial distribution of activity associated with each task. Functional-connectivity profiles showed stronger Real-Generated correspondence than Real-noise correspondence across the cortical network (Fig. 3b).

Participant-level cortical maps reproduced the principal positive and negative spatial patterns in lateral and medial views (Fig. 3c). The corresponding functional-connectivity matrices showed similar patterns of regional correlation in Real BOLD and Generated BOLD (Fig. 3d). These outputs allowed cortical mapping and functional-connectivity analysis to use the same generated signals. In the frequency-band analysis, SHAP values \cite{SHAP} were largest for gamma-conditioned Generated BOLD in visual-task decoding across all four conditions (Fig. 3e). The generated responses retained frequency-dependent features relevant to task discrimination.

\subsection*{The Haemoglobin Branch Generates Task-Evoked fNIRS Responses}

\begin{figure*}[tp]
\centering
\includegraphics[width=\linewidth,height=0.88\textheight,keepaspectratio]{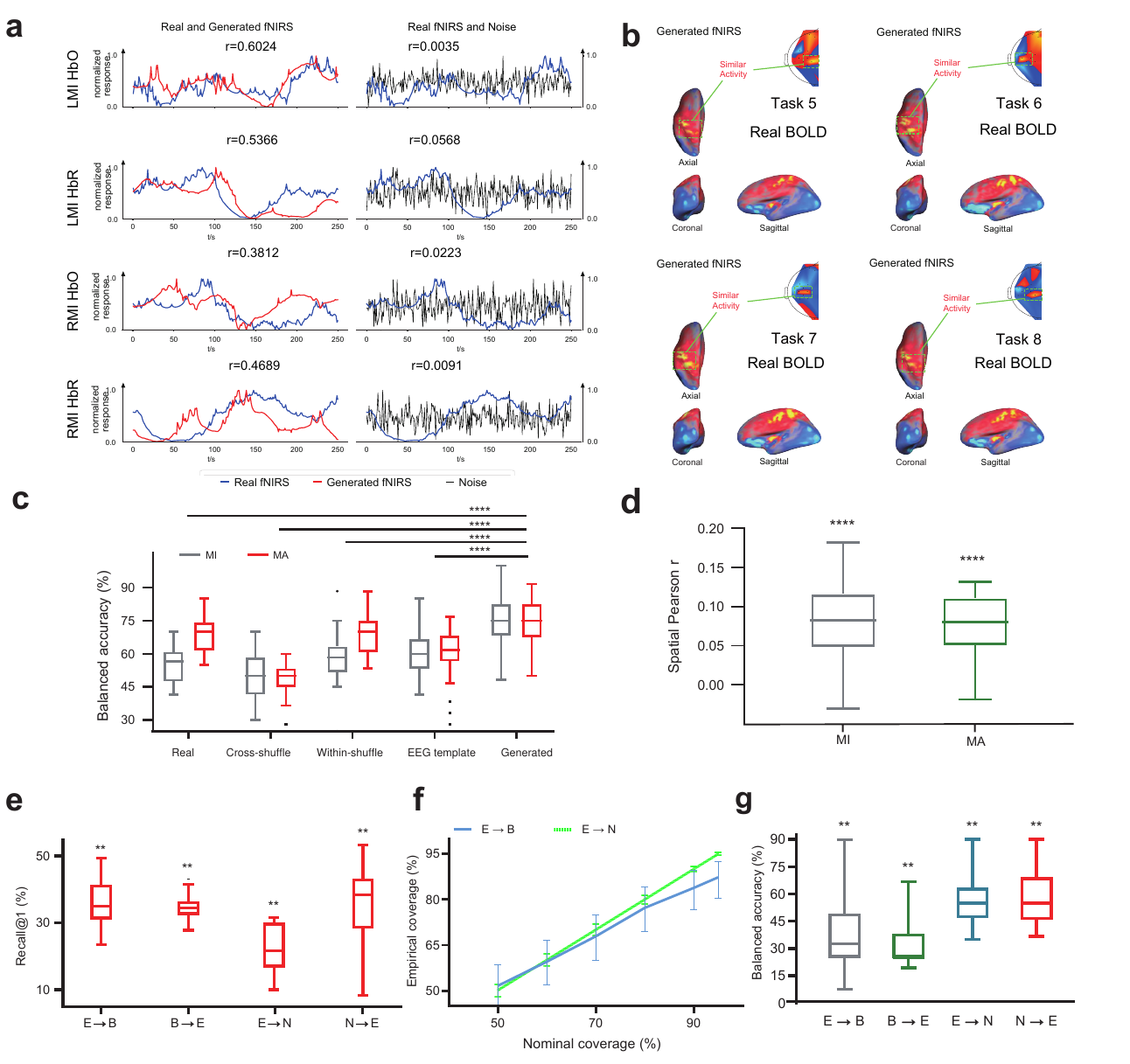}
\caption{\textbf{Haemoglobin generation supports task-evoked fNIRS analysis and paired target-domain use.} \textbf{a,} Real fNIRS haemoglobin waveforms and Generated fNIRS haemoglobin waveforms show response correspondence for left motor imagery and right motor imagery in HbO and HbR. Correlations between Real fNIRS and noise provide the reference control. \textbf{b,} Generated fNIRS haemoglobin response fields align with Real BOLD activation patterns across Tasks 5 to 8. \textbf{c,} Balanced accuracy compares Real, cross-shuffle, within-shuffle, EEG template and Generated conditions for motor imagery and mental arithmetic. \textbf{d,} Spatial Pearson $r$ quantifies motor-imagery and mental-arithmetic response-map correspondence. \textbf{e,} Recall at rank 1 evaluates paired retrieval between EEG features and the BOLD response space or the haemoglobin response space. \textbf{f,} Empirical coverage evaluates calibrated prediction intervals in the BOLD and haemoglobin response spaces. \textbf{g,} Balanced accuracy evaluates cross-domain decoding across the same feature and response spaces. E, B and N denote EEG, BOLD and fNIRS haemoglobin spaces. Boxes show median, interquartile range and 1.5-IQR whiskers. Paired response contrasts used two-sided participant-level sign-flip tests. Retrieval and directional decoding used one-sided condition-matched null tests. Holm correction was applied within each analysis family. Stars denote corrected significance levels with $^{*}P<0.05$, $^{**}P<0.01$, $^{***}P<0.001$ and $^{****}P<0.0001$.}
\label{fig4}
\end{figure*}

The haemoglobin target head generated task-evoked fNIRS waveforms. Generated HbO and HbR tracked their Real counterparts for both left and right motor imagery, with Pearson $r$ values above the Real-noise reference (Fig. 4a). Correlations ranged from 0.3812 to 0.6024 in the four displayed waveform comparisons. Generated fNIRS haemoglobin response fields also aligned with Real BOLD activation patterns across four visual tasks (Fig. 4b), extending the spatial correspondence to a second haemodynamic target domain.

Generated fNIRS haemoglobin features supported within-paradigm decoding for both motor imagery and mental arithmetic. Their balanced accuracy exceeded cross-shuffle, within-shuffle and EEG template controls in both paradigms (Fig. 4c). A classifier fitted on Real responses decoded the Generated responses, demonstrating that the haemoglobin branch retained task distinctions usable in the target domain. Spatial Pearson $r$ values were positive and significant for motor imagery and mental arithmetic, showing that Generated haemoglobin response maps preserved task-related spatial structure (Fig. 4d).

Recall at rank 1 demonstrated bidirectional paired retrieval between EEG features and either the BOLD or haemoglobin response space (Fig. 4e). Empirical coverage increased with nominal coverage in both target domains (Fig. 4f). Cross-domain decoding also identified task-related structure in all four directed comparisons (Fig. 4g). The retrieval and decoding results showed that the framework maintained useful correspondence between EEG features and both haemodynamic targets.

\subsection*{Generated BOLD Supports Decoding and Localization}

\begin{figure*}[!t]
\centering
\includegraphics[width=\linewidth,height=0.74\textheight,keepaspectratio]{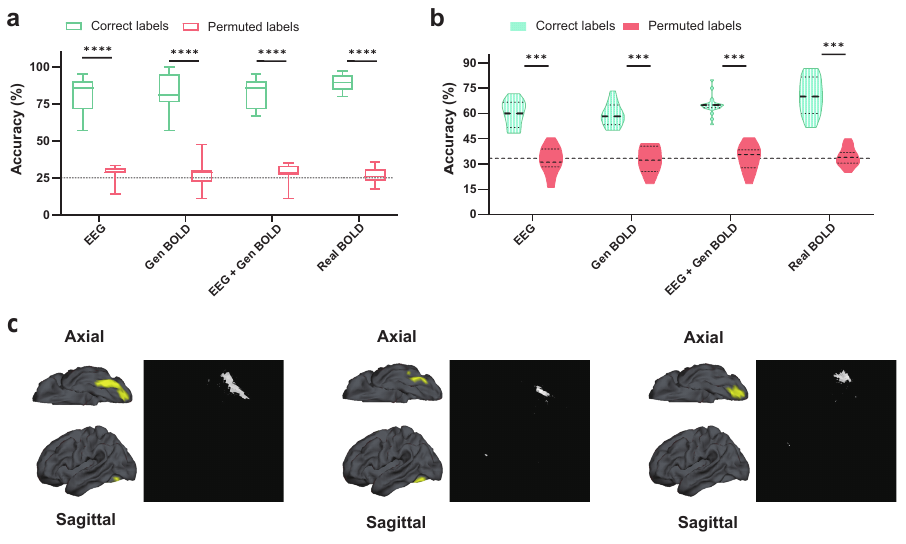}
\caption{\textbf{Generated target-domain features support calibrated decoding and localization.} \textbf{a,} Four-condition visual-task accuracy compares EEG features, Generated BOLD features, EEG features combined with Generated BOLD features and Real BOLD features under correct and permuted labels in 19 participants. The dotted line denotes 25\% chance. \textbf{b,} Three-condition independent-record accuracy compares the same feature sets under correct and permuted labels in 20 participants. The dotted line denotes 33.3\% chance. \textbf{c,} An external localization application displays representative BOLD response regions obtained by applying a NODDI-trained EEG-fMRI interface to Parkinson EEG inputs. Boxes show median, interquartile range and 1.5-IQR whiskers. Violins show participant distributions, with internal lines marking the median and quartiles. Correct-label and permuted-label contrasts used two-sided participant-level sign-flip tests with Holm correction. Stars denote corrected significance levels with $^{*}P<0.05$, $^{**}P<0.01$, $^{***}P<0.001$ and $^{****}P<0.0001$.}
\label{fig5}
\end{figure*}

In the four-condition visual-task analysis across 19 participants, Generated BOLD achieved high accuracy under correct labels and outperformed its permuted-label control (Fig. 5a). EEG features, Generated BOLD features, EEG combined with Generated BOLD and Real BOLD were evaluated under the same design, with 25\% chance as the four-condition reference. Correct-label distributions were separated from their permuted-label counterparts for all four feature sets. Generated BOLD alone retained discriminative task structure in the target domain and could also be combined with EEG features for decoding.

A distinct three-condition independent-record analysis across 20 participants used a 33.3\% chance reference and showed the same correct-label advantage (Fig. 5b). Generated BOLD and EEG combined with Generated BOLD retained accuracy above their permuted-label controls, while Real BOLD provided the target-domain benchmark. Task distinctions in the generated features remained accessible when a classifier fitted on one record was applied to a held-out record. The framework thus preserved useful BOLD features across the two decoding designs.

The external localization application used a NODDI-trained EEG-fMRI interface to place Parkinson EEG-derived BOLD response regions in anatomical context (Fig. 5c). This application extended the framework to anatomically localized BOLD generation from external EEG inputs.

\subsection*{Data Augmentation for Robust Brain Decoding}

\begin{figure*}[tp]
\centering
\includegraphics[width=\linewidth,height=0.86\textheight,keepaspectratio]{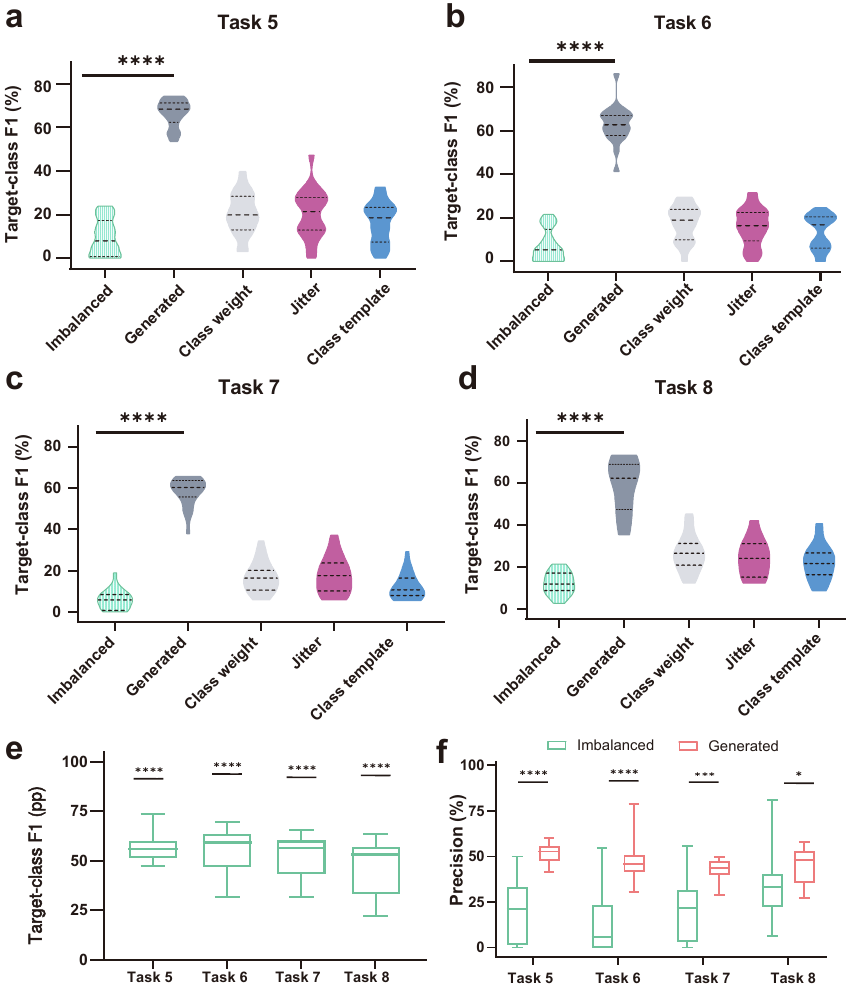}
\caption{\textbf{Generated target-domain augmentation improves low-sample brain decoding.} \textbf{a to d,} Target-class F1 is shown for Task 5, Task 6, Task 7 and Task 8, respectively, under imbalanced training, Generated augmentation, class weighting, jittering and class template augmentation. Each scenario starts from a low-sample target task and matched higher-sample non-target tasks. \textbf{e,} Generated minus imbalanced differences summarize target-class F1 gains in percentage points for Task 5 to Task 8 in the same order. \textbf{f,} Precision compares imbalanced and Generated training for each task. Violins show participant distributions, with dashed lines marking the median and quartiles. Boxes show median, interquartile range and 1.5-IQR whiskers. Generated and imbalanced conditions were compared with two-sided participant-level sign-flip tests and Holm correction. Stars denote corrected significance levels with $^{*}P<0.05$, $^{**}P<0.01$, $^{***}P<0.001$ and $^{****}P<0.0001$.}
\label{fig6}
\end{figure*}

The augmentation interface improved low-sample decoding. Each task defined a low-sample target condition, and Generated examples enriched that condition during decoder training. Generated augmentation increased target-class F1 for Task 5, Task 6, Task 7 and Task 8, exceeding imbalanced training with corrected significance in every task (Fig. 6a to e). Generated augmentation also produced higher median F1 than class weighting, jittering and class template augmentation across all four scenarios (Fig. 6a to d). The advantage over these reference methods showed that EEG-conditioned BOLD generation provided useful task-specific training examples.

Precision increased under Generated augmentation in all four tasks, accompanying the target-class F1 gains (Fig. 6f). The added examples improved decoding on held-out Real BOLD, linking the generation benefit to performance in the target domain. The framework used matched EEG inputs to enrich decoder training while keeping the available Real BOLD training set fixed.

\section*{Discussion}

EEG, fMRI and fNIRS rely on different acquisition principles. EEG records fast electrical potentials at the scalp, fMRI detects BOLD contrast through magnetic resonance, and fNIRS estimates HbO and HbR changes through optical absorption. The proposed framework aligns these paired recording streams in a unified representation space and generates BOLD or haemoglobin responses in their native target coordinates. The generated responses reproduced regional dynamics, cortical maps, functional connectivity and task-dependent structure used in functional neuroimaging analysis.

Generated BOLD preserved cortical response maps and functional connectivity, while Generated haemoglobin retained task-evoked waveform and spatial structure. These quantities extend the evidence beyond pointwise signal similarity because cortical maps, connectomes and haemoglobin response fields form the basis of many functional analyses \cite{Cortex-Align,fries2007gamma,natureplus1}. Unified representation aligns electrophysiological and haemodynamic features in compatible coordinates for generating target-specific BOLD and haemoglobin responses.

After paired calibration, additional EEG sessions can be mapped into target-domain representations compatible with cortical and network analysis. Paired multimodal acquisition becomes a calibration resource that supports functional analysis across subsequent recording sessions. This design broadens the analytical value of paired cohorts and supports missing-modality completion in task-rich or longitudinal studies.

The fNIRS target head applies the same representation principle to haemoglobin responses. Generated waveforms followed Real HbO and HbR dynamics, and Generated response fields aligned with Real BOLD activation patterns across tasks. Within-paradigm decoding showed that the haemoglobin outputs retained motor-imagery and mental-arithmetic structure. Bidirectional retrieval and cross-domain decoding further connected EEG features with BOLD and haemoglobin response spaces. A single representation framework generated target-specific responses for downstream analyses in both fMRI and fNIRS \cite{natureBCI,nmi9,nbe1}.

Generated BOLD and haemoglobin responses supported decoding across visual, motor-imagery and mental-arithmetic paradigms. Correct-label and permuted-label contrasts identified task-relevant structure in the Generated BOLD features, while the external application placed EEG-derived BOLD responses in anatomical context. The framework supported task decoding in paired cohorts and anatomical localization from external EEG inputs.

The unified representation interface combines targeted multimodal calibration with denser electrophysiological sampling. Paired recordings define the multimodal correspondence, and subsequent EEG recordings extend target-domain analysis across additional sessions or conditions. This design is relevant to longitudinal functional monitoring, repeated cognitive assessment, adaptive brain-computer interfaces and individualized multimodal brain profiling, where dense electrophysiological sampling is more feasible than repeated multimodal acquisition.

Low-sample augmentation provides a direct application of the generated target domain. Generated BOLD examples improved target-class F1 under sparse training, and the precision gains confirmed higher positive predictive performance. This result addresses brain-computer interfaces and neuroimaging studies in which target-domain observations are unevenly distributed across tasks, participants or visits \cite{nbe2,nm2,nc3}. By enriching scarce target conditions, the framework improves decoding on held-out Real BOLD and supports data augmentation for robust brain decoding.

Recent wearable fNIRS and clinical EEG studies emphasize dense sampling, individualized prediction and automated detection \cite{Jounghani2025WearablefNIRS,Sun2024AdaptiveEEG,Li2026HeroinEEGFC,Wang2026MSAEEG,Sun2025VIPEEGNet}. In this context, the proposed framework combines dense EEG sampling with targeted haemodynamic calibration for longitudinal monitoring, task-rich brain decoding and cross-subject modelling \cite{nmi13,nmi17,nm1,nm3}. The method converts paired multimodal recordings into a reusable model for generating BOLD and haemoglobin responses from EEG.

\section*{Methods}

\subsection*{Datasets and Evaluation Scope}

This study uses four public datasets with distinct roles. NatView and the EEG-fNIRS motor-feedback cohort provide paired EEG-fMRI and EEG-fNIRS cohorts for direct evaluation. NODDI provides an independent simultaneous EEG-fMRI resource for fitting the EEG-to-BOLD interface used in the external localization application. EEG-3SAORP provides Parkinson EEG inputs for that application. The quantitative results focus on paired cohorts with Real target-domain references, and the external localization panel provides anatomical context for clinical EEG inputs.

The BOLD analyses evaluate reconstruction, cortical response mapping, task decoding and data augmentation. The haemoglobin analyses evaluate HbO and HbR waveforms, event-locked response features and within-paradigm decoding. Retrieval, prediction intervals and directional decoding in Fig. 4e to g evaluate calibrated correspondence between EEG features and haemodynamic response spaces.

NatView provides simultaneous EEG-fMRI recordings during naturalistic viewing \cite{dataset_1}. The primary BOLD implementation uses 21 participants, 141 runs and 15,510 paired frames. A locked quality-controlled analysis subset contains 12,345 frames with valid EEG histories and BOLD targets. The task labels serve as stimulus identifiers. Session 2 uses Task 1 (Despicable Me Hungarian), Task 2 (Inscapes), Task 3 (Monkey2) and Task 4 (Monkey5). Session 1 uses Task 5 (Despicable Me English), Task 6 (Inscapes), Task 7 (Monkey1) and Task 8 (The Present). The two Inscapes recordings have different session labels. Five participant-disjoint outer folds assign generator fitting, tuning and evaluation roles. Fitting data determine scaling, imputation and target transformations. Tuning data guide model selection. Outer-test labels are reserved for final evaluation.

The EEG-fNIRS motor-feedback dataset contains 29 participants \cite{dataset_2}. The task-evoked response analysis uses 1,740 motor-imagery events and 1,740 mental-arithmetic events, with 36 HbO and 36 HbR channels. Five participant-disjoint folds assign 18 to 19 fitting participants, five tuning participants and five to six evaluation participants. EEG and fNIRS event clocks are aligned using their respective markers. Response analyses use EEG features and haemoglobin concentrations in native physical coordinates.

The NODDI dataset contains simultaneous EEG and fMRI recordings from healthy adults \cite{noddi1_deligianni2016noddi,noddi2_deligianni2014relating}. The external localization application uses this dataset to fit an EEG-to-BOLD interface before applying the fitted interface to Parkinson EEG inputs. The EEG-3SAORP Parkinson dataset contains resting-state and three-stimulus auditory oddball EEG recordings \cite{Parkinson_cavanagh2021eeg}. EEG-3SAORP provides the external EEG inputs for the Parkinson localization display in Fig. 5c.

The primary quantitative analyses use the paired EEG-fMRI and EEG-fNIRS cohorts described above. NODDI and EEG-3SAORP jointly define the external localization application in Fig. 5c and provide anatomical context outside the paired reconstruction analyses.

\subsection*{Ethics and Original Acquisition}

This study analyses previously acquired neuroimaging data from de-identified public releases. The original NatView study received approval from the Nathan Kline Institute Institutional Review Board (IRB\#941632), with written informed consent before acquisition \cite{dataset_1}. The original EEG-fNIRS motor-feedback acquisition was approved by the Ethics Committee of the Institute of Psychology and Ergonomics, Technical University of Berlin (approval number SH 01 20150330), and participants provided informed consent before the experiment \cite{dataset_2}. NODDI and EEG-3SAORP were analysed as public de-identified data releases \cite{noddi1_deligianni2016noddi,noddi2_deligianni2014relating,Parkinson_cavanagh2021eeg}.

\subsection*{Preprocessing and Signal Quality}

Original preprocessing included EEG filtering and artefact removal, FreeSurfer fMRI registration and cortical sampling \cite{FreeSurfer,nature4}, and optical conversion to filtered HbO/HbR. The six-second alignment convention follows the original processed interface \cite{six_seconds}. Analyses using physical signals retain recorded clocks and event definitions.

For the BOLD analysis summarized in Fig. 2, corrected EEG features include scanner and cardiac artefact correction. The locked signal mask retains 12,345 frames with valid BOLD targets and EEG histories across all 21 participants and 141 runs. Missing feature values are imputed from fitting participants. Anatomical target values are the means of 142 bilateral Destrieux regions. Run-centred BOLD targets and whole-run EEG context define the run-level reconstruction analysis. Regions, participants and display runs follow anatomical and protocol definitions.

The event-locked haemoglobin analysis in Fig. 4c,d uses 30 common-average-referenced EEG scalp channels filtered in 8 to 13, 13 to 30 and 8 to 30 Hz bands. EEG power and covariance features summarize 0 to 10 s after each event. Haemoglobin targets are baseline-corrected relative to \(-3\) to \(0\) s and contain 5 to 15 s response means and slopes in native concentration coordinates. This response-feature analysis includes all 3,480 events. The waveform comparisons in Fig. 4a use temporal HbO and HbR outputs, displayed as normalized responses.

\subsection*{Generative Framework and Unified Representations}

The framework uses EEG as the electrophysiological input and generates fMRI BOLD and fNIRS haemoglobin responses through a unified representation formulation. The workflow in Fig. 1 combines feature extraction, modal patch embedding, conjugate alignment, generation and modal unpatching. Pre-trained EEG and fMRI feature models provide representation components \cite{LaBraM,UKB_Sudlow2015UKBA}. Source transformations and target mappings are calibrated for each cohort and haemodynamic domain. Let \(m\in\{B,H\}\) denote the BOLD and haemoglobin domains. For participant \(i\), run or session \(r\), and time or event index \(t\), the paired fitting set is

\begin{equation}
\begin{aligned}
\mathcal D_m
&=
\Bigl\{\bigl(
\mathbf e_{ir}(t-\Delta_m:t),\\
&\qquad
\mathbf y^m_{irt},
\mathbf q_{rt}
\bigr)\Bigr\}_{i,r,t}
\end{aligned}
\label{eq_paired_set}
\end{equation}

\(\mathbf e_{ir}(t-\Delta_m:t)\) is the EEG history, \(\mathbf y^m_{irt}\in\mathbb R^{d_m}\) is the target-domain vector, and \(\mathbf q_{rt}\) records acquisition context. The BOLD domain uses cortical targets, and the haemoglobin domain uses HbO and HbR targets. Target-specific patch operators convert encoded source and target streams into token sets,

\begin{equation}
\begin{aligned}
\mathbf r^E_{irt,q}
&=
E^E_q\Pi^E_q\!\left(
\mathbf f^E_{irt}
\right)+\mathbf b^E_q\\
\mathbf r^m_{irt,q}
&=
E^m_q\Pi^m_q\!\left(
\mathbf f^m_{irt}
\right)+\mathbf b^m_q
\end{aligned}
\label{eq_modal_patch}
\end{equation}

\(\Pi^E_q\) and \(\Pi^m_q\) select source and target patches, \(E^E_q\) and \(E^m_q\) embed these patches, and \(q\) indexes temporal, spatial or channel-wise tokens. This operation preserves modality-specific sampling structure before the source and target streams enter the shared representation interface.

Source and target feature maps encode the paired streams before target-domain generation,

\begin{equation}
\begin{aligned}
\mathbf f^E_{irt}
&=
\phi_{E,m}\!\left(
\mathbf e_{ir}(t-\Delta_m:t)
\right)\\
\mathbf f^m_{irt}
&=
\phi_m\!\left(
\mathbf y^m_{irt}
\right)
\end{aligned}
\label{eq_encoders}
\end{equation}

\(\phi_{E,m}\) and \(\phi_m\) denote the source and target feature maps for domain \(m\). Their fitting-participant transformations define the corresponding feature coordinates. Hyperdimensional integration projects source and target features into compatible coordinates,

\begin{equation}
\begin{aligned}
\mathbf g^E_{irt}
&=
U^E_m\!\left[
\mathbf f^E_{irt},
\{\mathbf r^E_{irt,q}\}_{q=1}^{Q_E}
\right]\\
\mathbf g^m_{irt}
&=
U^m\!\left[
\mathbf f^m_{irt},
\{\mathbf r^m_{irt,q}\}_{q=1}^{Q_m}
\right]
\end{aligned}
\label{eq_conjugate_projection}
\end{equation}

\(U^E_m\) and \(U^m\) are target-aware projections that align electrophysiological and haemodynamic features. The conjugate representation provides the paired coordinates for source-conditioned target generation.

Modal unpatching restores generated target codes to the coordinates used for cortical or haemoglobin analysis,

\begin{equation}
\begin{aligned}
\hat{\mathbf y}^m_{irt}
&=
\mathcal U_m\!\left(\hat{\mathbf u}^m_{irt}\right)
\end{aligned}
\label{eq_modal_unpatch}
\end{equation}

\(\mathcal U_m\) denotes the target-specific inverse transformation. It restores cortical values for BOLD and haemoglobin response coordinates for fNIRS using fitting-participant scaling and spatial definitions.

The conditional BOLD model integrates EEG history through learnable temporal weights,

\begin{equation}
\begin{aligned}
\alpha_{\ell q}
&=
\frac{
\exp(\omega_{\ell q})
}{
\sum_v\exp(\omega_{\ell v})
}\\
\mathbf h_{irt,\ell}
&=
\sum_q\alpha_{\ell q}\tilde{\mathbf r}^E_{irt,q}
\end{aligned}
\label{eq_temporal_kernel}
\end{equation}

\(\omega_{\ell q}\) is a learned weight for temporal profile \(\ell\) and token \(q\), and \(\tilde{\mathbf r}^E_{irt,q}\) is the temporally encoded EEG token. The paired BOLD interface uses the six-second alignment convention. Temporal weighting adapts the integration of the EEG history within this interface. The event-locked haemoglobin analysis uses the response windows specified above.

Source-conditioned and paired target representations provide the coordinates for target estimation,

\begin{equation}
\begin{aligned}
\mathbf z^m_{irt}
&=
\Psi_m\!\left(
\mathbf g^E_{irt},
\mathbf q_{rt}
\right)\\
\mathbf u^m_{irt}
&=
P_m\mathbf f^m_{irt}
\end{aligned}
\label{eq_unified_code}
\end{equation}

\(\mathbf z^m_{irt}\) is the source-conditioned representation, and \(\mathbf u^m_{irt}\) is the paired target code. In the conditional BOLD model, attention queries integrate the weighted temporal context and EEG tokens to form activity and context representations. In the haemoglobin response model, a pooled EEG projection aligns the two paradigms with their haemoglobin targets. Target-specific mappings preserve the coordinates required for the corresponding downstream analyses.

The conditional BOLD model estimates a distribution in the fitted target space,

\begin{equation}
\begin{aligned}
&p_{\theta_B}\!\left(
\mathbf u^B_{irt}\mid
\mathbf e_{ir}(t-\Delta_B:t)
\right)\\
&\quad =
\mathcal N\!\left(
\boldsymbol\mu^B_{\theta,irt},
\Sigma^B_{\theta,irt}
\right)
\end{aligned}
\label{eq_conditional_distribution}
\end{equation}

The conditional mean combines a fitted linear anchor with a learned attention residual,

\begin{equation}
\begin{aligned}
\hat{\mathbf u}^B_{irt}
&=
\boldsymbol\mu^B_{\theta,irt}\\
&=
W^B_{\mathrm{anchor}}\mathbf f^E_{irt}
+b^B_{\mathrm{anchor}}\\
&\quad
+R_{\theta_B}\!\left(
\mathbf z^B_{irt}
\right)
\end{aligned}
\label{eq_target_decoder}
\end{equation}

\(R_{\theta_B}\) uses cross-attention features together with EEG-derived activity and context codes. The fitted anchor and residual are combined in the BOLD target space before inverse projection. The haemoglobin model uses a pooled source projection and paradigm-specific regularized response maps, as specified below.

The conditional BOLD model is fitted with target likelihood, paired representation alignment and activity-context constraints,

\begin{equation}
\begin{aligned}
\mathcal L_B
&=
\mathcal L_{\mathrm{NLL}}
+\lambda_a\mathcal L_{\mathrm{align}}\\
&\quad
+\lambda_c\bigl(
\mathcal L_{\mathrm{cond}}^E
+\mathcal L_{\mathrm{cond}}^B
\bigr)\\
&\quad
+\lambda_o\mathcal L_{\mathrm{orth}}
\end{aligned}
\label{eq_training_loss}
\end{equation}

\(\mathcal L_{\mathrm{NLL}}\) is the Gaussian negative log likelihood, and \(\mathcal L_{\mathrm{align}}\) is a symmetric contrastive objective for paired activity codes. Fitting-derived condition codes supervise EEG and target activity representations through \(\mathcal L_{\mathrm{cond}}^E\) and \(\mathcal L_{\mathrm{cond}}^B\). \(\mathcal L_{\mathrm{orth}}\) penalizes the squared inner product between normalized activity and context codes. Weights \(\lambda_a\), \(\lambda_c\) and \(\lambda_o\) are 0.10, 0.20 and 0.01, respectively. Transformations are estimated from fitting participants. Tuning-participant reconstruction error selects saved model states, and tuning data determine downstream hyperparameters before evaluation participants are scored.

\subsection*{Reconstruction Interface}

The primary BOLD generator uses 21 participants, 141 runs and 15,510 frames, with 110 consecutive frames per run. Native EEG has 61 channels at 250 Hz. Eight temporal bins and six spectral groups supply log-power features. The analysis uses the released preprocessed interface. Fit-participant transformations define target-domain coordinates. Real BOLD targets are centred using the low-motion mean of each run. Generated responses are evaluated in these centred coordinates. Participant-disjoint folds evaluate every participant once.

Generated outputs are projected to cortical values. All 142 bilateral Destrieux regions \cite{destrieux2010automatic} enter reconstruction. The observed run-demeaned target is the reference. A fixed within-run cyclic input shift serves as the temporal control. Temporal correlations are computed within each run and region. Runs are averaged within task, and tasks are averaged equally within participant. Temporal SD summaries compare the Generated scale with the Real scale directly. Inference receives EEG features only.

\subsection*{Organizational Correspondence and Specificity}

Cortical analyses use common anatomical coordinates for Real BOLD and Generated BOLD. Task-wise key-region maps and regional connectivity profiles quantify spatial and network correspondence in Fig. 3a,b. The participant-level maps and matrices in Fig. 3c,d retain the cortical sampling and 71-region connectivity ordering of the corresponding analysis. Functional connectivity is calculated from temporal correlations between regional signals. Frequency-band attribution in Fig. 3e evaluates visual-task decoding from Generated BOLD conditioned separately on delta, theta, alpha, beta and gamma EEG bands. SHAP values \cite{SHAP} are calculated for the task-decoding model and summarized by frequency band.

\begin{equation}
\begin{aligned}
C^s_{ab}(i,r)
&=
\operatorname{corr}_t\!\left(
Y^s_{ir,a}(t),
Y^s_{ir,b}(t)
\right)\\
r_{\mathrm{FC}}(i,r)
&=
\operatorname{corr}_{a<b}\!\left(
C^{\mathrm{Real}}_{ab}(i,r),
\right.\\
&\qquad\left.
C^{\mathrm{Generated}}_{ab}(i,r)
\right)
\end{aligned}
\label{eq_functional_connectivity}
\end{equation}

\(s\in\{\mathrm{Real},\mathrm{Generated}\}\) identifies the signal type, and \(a,b\) index cortical regions.

\subsection*{Joint Haemoglobin Response Model and Locked Decoding}

A pooled fit-only EEG transformation is shared across motor imagery and mental arithmetic. The haemoglobin target head predicts a 144-coordinate response using an acquisition-context code that identifies the recording protocol without encoding the within-paradigm class label. Downstream classifiers evaluate held-out within-paradigm labels under the locked decoding protocol. Fitting and tuning participants determine model selection. A classifier fitted on Real responses is locked before Generated responses from outer-test participants are scored.

\begin{equation}
\begin{aligned}
\hat{\mathbf h}_{irt}
&=
\left[
\hat{\mathbf h}^{\mathrm{HbO}}_{irt},
\hat{\mathbf h}^{\mathrm{HbR}}_{irt}
\right]\\
&=
G_{\theta_H}\!\left(
\mathbf f^E_{irt},
\mathbf q_{rt}
\right)
\end{aligned}
\label{eq_hb_response}
\end{equation}

Across-class shuffling permutes fitting donors within participant, paradigm and session. Within-class shuffling additionally preserves the within-paradigm label. EEG-selected templates use fitting-participant Real response means, with template selection based on EEG. Tuning participants guide model selection. Real responses and the fit mean provide references. Spatial contrasts between the two conditions are formed separately for Real and Generated responses, and their HbO and HbR correlations are averaged within participant.

\subsection*{BOLD Target Head for Decoding and Augmentation}

The BOLD target head used for decoding and data augmentation follows Eq.~\eqref{eq_target_decoder}. Five participant-disjoint folds fit the generator. Fit-only EEG log-power features and fit-only BOLD target coordinates define the paired representation interface. The model uses a hidden width of 128, four attention layers, eight heads and 64-dimensional activity and context codes. Four fitting-derived activity states supervise the representation objective. Context representations are inferred from EEG features. Visual-stimulus labels are used by the downstream decoding analysis. AdamW fitting uses a learning rate of 0.0003, weight decay of 0.0002, batches of 1,024 and a maximum of 12 epochs with patience 3. Tuning-participant reconstruction error selects the saved state before downstream scoring.

The decoding and augmentation experiments use conditional-mean responses from this target head. Short-window mean and SD summaries in the fitted target coordinate system yield target-domain features. Transformations, fold roles, selected weights and downstream windows are fixed before evaluation.

\subsection*{Calibrated BOLD Task Decoding}

The BOLD target-head decoding analysis in Fig. 5a uses all 19 participants with complete four-condition visual-task data. The four labels define a 25\% chance reference. Each participant receives generated responses from a participant-disjoint generator fold. Each run is divided into non-overlapping windows, with dedicated fitting, tuning and evaluation windows. Four feature sets share this role structure. The feature sets are EEG features, Generated BOLD features, Real BOLD features and EEG features combined with Generated BOLD features. Every condition uses the same fit-only scaling, regularized classifier and tuning-guided model selection. Classifier type and search budget are matched across feature sets.

\begin{equation}
\begin{aligned}
\hat{k}
&=
\arg\max_k
P_{\eta}\!\left(k\mid\mathbf v\right)\\
\mathbf v
&\in
\bigl\{
\mathbf v_E,
\mathbf v_{\mathrm{Generated}},\\
&\quad
\mathbf v_E\oplus\mathbf v_{\mathrm{Generated}},
\mathbf v_{\mathrm{Real}}
\bigr\}
\end{aligned}
\label{eq_decoding_rule}
\end{equation}

Fitting-label permutations preserve within-participant class counts. Only fitting labels are permuted. Tuning labels guide model selection, and test labels are used for final scoring. Participant roles, window roles and null scoring are set before evaluation.

\subsection*{Independent-Record Decoding}

The independent-record test in Fig. 5b uses 20 participants with complete three-condition records. The three labels define a 33.3\% chance reference. The analysis fits one record and applies the trained classifier to the held-out record under the same feature-set definitions used in Fig. 5a. The fitting-label permutation leaves tuning and test labels unchanged. Accuracy is the displayed endpoint. Correct-label and permuted-label analyses use matched participant roles, feature scaling, classifier selection and null scoring.

\subsection*{Cross-Domain Calibration, Retrieval and Decoding}

The paired feature-space analyses in Fig. 4e to g use analysis-specific calibration. Paired retrieval selects an alignment on tuning participants, ranks eight condition-matched candidates across EEG and haemodynamic response spaces, and reports recall at rank 1 in both directions. The prediction-interval analysis fits an EEG-to-target map and uses calibration-residual quantiles to define intervals at each nominal coverage. Empirical coverage is evaluated on held-out target responses. Directional decoding uses calibrated source-target features and a locked target-domain classifier, reporting balanced accuracy for the four directed pairs in Fig. 4g.

\subsection*{Data Augmentation for Robust Brain Decoding}

Fig. 6 uses a locked generator in five participant-disjoint folds. Fig. 6a to d present the low-sample target tasks in ascending task order from Task 5 to Task 8. Each example summarizes a short target-domain window with mean and variability features. Each task becomes the low-sample target in a task-specific scenario, with 30 Real target examples and 120 Real examples per other task available to the downstream decoder. Generation adds 90 model-generated target-domain examples from matched EEG inputs without increasing the number of Real BOLD examples used for decoder fitting. Jitter and class template augmentation add the same count. Class weighting uses the original Real examples. Regularization and augmentation parameters are chosen by inner validation.

\begin{equation}
\begin{aligned}
\mathcal T^{\mathrm{aug}}_k
&=
\mathcal T^{\mathrm{Real}}_k
\cup
\left\{
\left(\mathbf v_{\mathrm{Generated},j},k\right)
\right\}_{j=1}^{90}\\
\mathbf v_{\mathrm{Generated},j}
&=
\mathcal S\!\left(
\{\boldsymbol\mu^B_{\theta,t}\}_{t\in\mathcal W_j}
\right)
\end{aligned}
\label{eq_augmentation_set}
\end{equation}

\(\mathcal S\) concatenates the mean and standard deviation over window \(\mathcal W_j\) in the fitted BOLD coordinates.

Testing uses Real BOLD from held-out participants. Paired calibration and downstream decoding use participant-disjoint roles, and the low-sample constraint is applied to the Real target-domain examples available for decoder fitting. Matched EEG inputs provide the source for Generated augmentation. Held-out Real BOLD is used only for evaluation. Sampling repetitions are averaged within participant. Low-sample F1, recall and precision are computed for participants with the target task present.

The augmentation analysis compares Generated examples with imbalanced training, class weighting, jittering and class template augmentation under the same fit-only scaling, validation search and held-out testing. The displayed Generated minus imbalanced F1 contrasts are evaluated with participant-level paired tests.

\subsection*{Statistical Analysis}

Participants are weighted equally. Sessions, trials, runs, regions and frames are treated as nested observations and aggregated within participant before participant-level inference. All eligible participants are retained, and uncertainty estimates preserve the saved fold assignments and model-selection decisions.

Planned comparisons are corrected within each analysis family. BOLD reconstruction, visual-task decoding, independent-record decoding, haemoglobin response analysis, cross-domain analysis and augmentation use their corresponding correction families.

Two-sided participant-level sign-flip tests evaluate the paired response, decoding and augmentation contrasts. Paired retrieval and directional cross-domain decoding use one-sided condition-matched null tests based on 1,000 permutations. Holm correction controls multiplicity within each analysis family. Response, decoding and augmentation endpoints are accompanied by the descriptive correlation, recall and precision summaries defined for each experiment.

Distribution displays use box and violin summaries with all finite observations included in descriptive calculations and statistical tests. Axis bounds contain the complete source ranges and relevant chance or null references. Fig.~2g uses a labelled logarithmic ordinate to accommodate the positive temporal SD range.

Throughout the figures, Generated denotes a model's target-domain output, Real denotes acquired target-domain observations, and EEG combined with Generated denotes concatenated EEG features and generated-target features. Target coordinates and output heads remain specific to each analysis. Shift, Phase and Shuffle denote cyclic temporal shifting, phase randomization and shuffled source-target pairing. Cross-shuffle and Within-shuffle denote the haemoglobin controls, and Permuted labels denotes a decoding model fitted after label permutation.


%
%
%
%
%
%
%
%
%
%

\setlength{\bibsep}{0.35em}
\bibliography{sn-bibliography}

\end{document}